\documentclass[aps,prb,preprint,superscriptaddress]{revtex4-1}
\usepackage{graphicx}
\usepackage{bm}
\begin{document}
\title{Algebraic decay of the nonadiabaticity arising through chiral spin transfer torque in magnetic domain walls with Rashba spin-orbit interaction}
\author{D. Wang}
\email{wangdaowei@sztu.edu.cn}
\affiliation{College of Engineering Physics, Shenzhen Technology University, Guangdong 518118, P. R. China}
\author{Yan Zhou}
\email{zhouyan@cuhk.edu.cn}
\affiliation{School of Science and Engineering, The Chinese University of Hong Kong, Shenzhen, Guangdong 518172, P. R. China}
\begin{abstract}
Spin transfer torque in a two dimensional electron gas system without space inversion symmetry was theoretically investigated by solving the Pauli-Schr\"{o}dinger equation for the itinerant electrons inside magnetic domain walls. Due to the presence of the Rashba spin-orbit coupling induced by the broken inversion symmetry, the spin transfer torque is chiral and the nonadiabaticity, which is defined to measure the relative importance of the nonadiabatic, field-like torque to the adiabatic, damping-like torque, exhibits an inverse power law decay as the domain wall width is increased. This algebraic decay is much slower than the exponential decay observed for systems without the Rashba spin-orbit coupling, and may find applications in innovative design of spintronic devices utilising magnetic topological textures such as magnetic domain walls and skyrmions.
\end{abstract}
\date{\today}
\maketitle
\section{Introduction}
From ancient times, the conventional way to manipulate a ferromagnet's magnetic state is through application of an external magnetic field. The situation changed drastically in the last two decades, following the innovative proposition by Berger \cite{Berger} and Slonczewski \cite{Slonczewski96} of using high density electric current to exert a torque, dubbed the spin transfer torque (STT), on local magnetization. After more than twenty years of intensive quest for unconventional methods for magnetization manipulation with high energy efficiency and operation speed, we now have many alternatives at our disposal, such as ultrashort laser pulses \cite{Kirilyuk}, electric field \cite{Ohno} and even magneto-elastic waves \cite{Dreher12}. Even with so many competitors on the arena for the manipulation of magnetization, electric current based methods, including the original STT, and the later developments of the Rashba spin-orbit torque (RSOT) \cite{Obata,Garate,Manchon,Matos-Abiague,Wang12} and spin-Hall effect \cite{she} in systems with spin-orbit interaction (SOI), attract more attention due to their easy implementation and compatibility with current semiconductor technology.

Investigations \cite{Bazaliy98,Li04,Thiaville05,Xiao06} following Berger's and Slonczewski's seminal works showed that the STT should be a sum of two terms, one damping-like torque which is already given in their original papers, and the other an additional field-like torque,
\begin{equation}
\bm {\tau} = \alpha\, \hat {j} \cdot \nabla \hat{M} + \beta\, \hat {M} \times (\hat {j} \cdot \nabla \hat{M}),
\label{stt}
\end{equation}
for a continuous distribution of magnetization characterized by the normalized magnetization vector, $\hat {M} = \textbf{M}/M$, where $\textbf{M}$ is the magnetization vector and $M$ the saturation magnetization. $\hat {j}$ is a unit vector pointing to the direction of the electric current flowing in a ferromagnet. $\alpha$ and $\beta$ are decomposition coefficients. As shown by a subsequent model quantum-mechanical investigation on STT inside a magnetic domain wall (DW) \cite{Xiao06}, the damping-like torque is given by the continuous rotation of the itinerant electron spin towards the local magnetization, while the field-like torque is attributable to the misalignment between the electron spin and the local magnetization inside the rigid DW. With this microscopic interpretation, the damping-like torque (first term in Eq. (\ref{stt})) and the field-like torque (second term in Eq. (\ref{stt})) are usually called the adiabatic torque and the nonadiabatic torque, respectively.

In systems without spatial inversion symmetry, such as the interfaces between ferromagnets and heavy metals, there could exist Rashba SOI along the symmetry-breaking direction \cite{Gambardella}. The mutual influence of the electron's spin and orbital dynamics induced by the Rashba SOI is derived from the fact that the Rashba SOI can be absorbed into the mechanical momentum operator, forming a covariant canonical momentum operator \cite{kim13}: The emergent gauge potential appearing in the covariant momentum operator depends on the spin operator of the electron, so the covariant momentum operator becomes chiral on the electron's motion. The motion of electrons under the influence of this chiral momentum operator is different from that of decoupled spin and orbital motion. Actually, due to the Rashba coupling between the spin and orbital degrees of freedom, the nonadiabatic RSOT inside magnetic DWs exhibits a topological behavior \cite{Wang20} that is related to the topology of the underlying DWs in the adiabatic limit. In addition, there are both experimental investigations and theoretical demonstrations on chiral gyromagnetic and Gilbert damping constants \cite{Jue,Akosa,Freimuth17} in magnetic DWs. Along the same line, we expect that the STT should also inherit to some extent the intrinsic chiral characteristic of the Rashba SOI. This expectation is actually borne out by our numerical results; both $\alpha$ and $\beta$ in Eq. (\ref{stt}) are chiral on the magnetization rotation in DWs. This chiral STT induced by the Rashba SOI can be utilized to design spintronic devices based on magnetic topological textures.

Another related but distinct feature bestowed on STT in DWs by the Rashba SOI is the algebraic decay of the STT nonadiabaticity. The STT nonadiabaticity, which is defined as the ratio $\beta/\alpha$, measures the relative importance of the nonadiabatic STT. It is an important parameter to consider for applications of STT in current-driven DW motion, as only the nonadiabatic STT can sustain a steady DW motion. It also plays an important role in the understanding of the origin of the STT, simply because it reveals the scattering characteristics of electrons in DWs. Previous investigation showed that it decays exponentially as the DW width is increased \cite{Xiao06}. This feature of the STT renders the utilization of the nonadiabatic STT component to drive DW motion in materials with sizable DW width difficult. However, we would like to emphasize that this exponential decay is obtained without considering the effect of any SOI. We will show in the following that, by including the Rashba SOI, the exponential decay of STT nonadiabaticity is reduced to an algebraic one, thus facilitating the utilization of the nonadiabatic STT in a wider range of material systems with SOI. The algebraic decay of the STT nonadiabaticity is the main result of our numerical investigation on STT in magnetic DWs under the influence of the Rashba SOI.

The organization of the paper is as follows. In Sec. \ref{theory} a brief description of the semiclassical theoretical framework is given. Our numerical results are presented in Sec. \ref{results}, starting from the discussion about the STT nonadiabaticity and then confirming the chiral character of the decomposition coefficients $\alpha$ and $\beta$. Finally, Sec. \ref{conclusion} summarizes our results on STT inside magnetic DWs with Rashba SOI.
\begin{figure}\centering
\begin{minipage}[c]{0.45\linewidth}
\includegraphics[width=\linewidth]{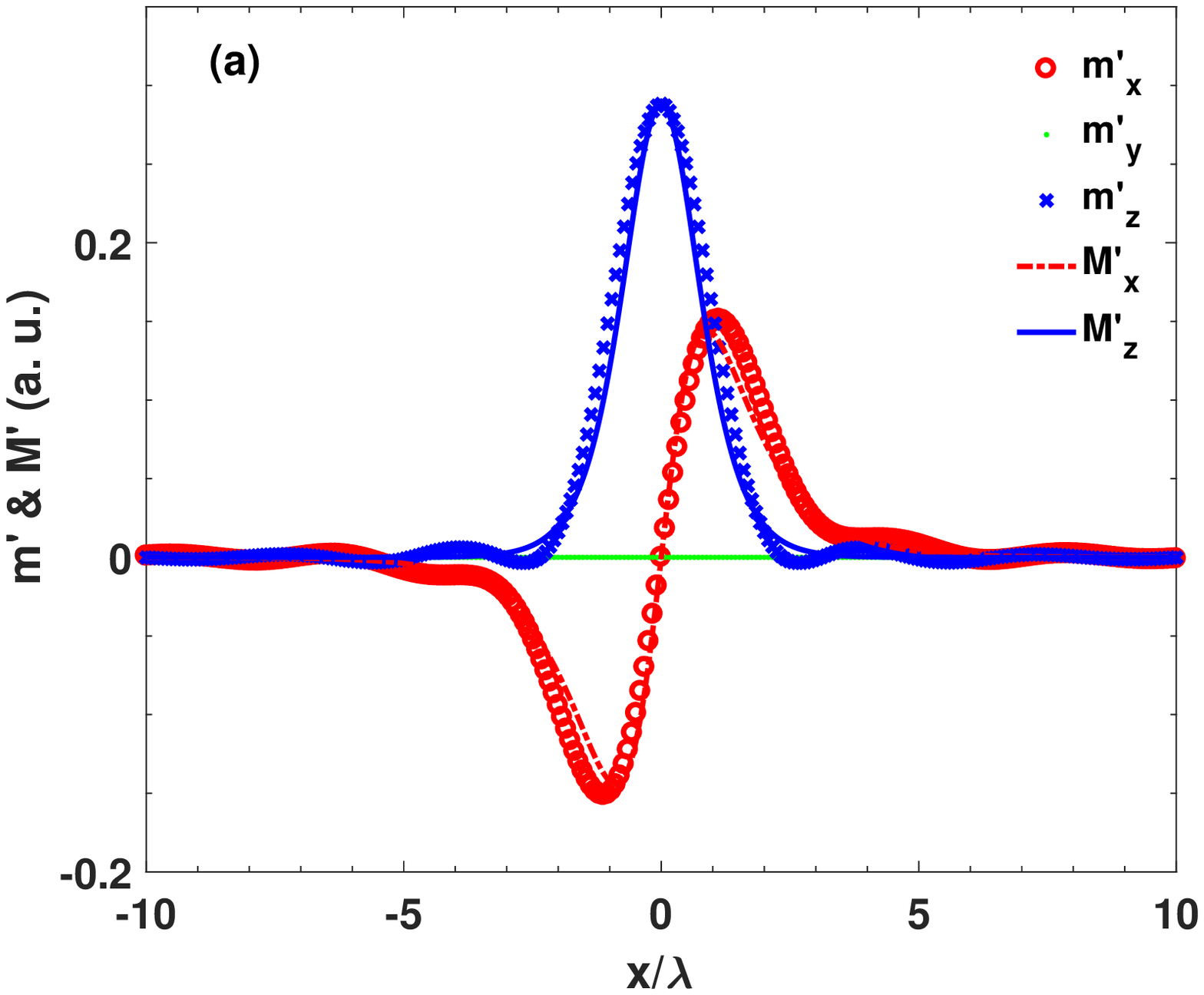}
\end{minipage}
\hfill
\begin{minipage}[c]{0.45\linewidth}
\includegraphics[width=\linewidth]{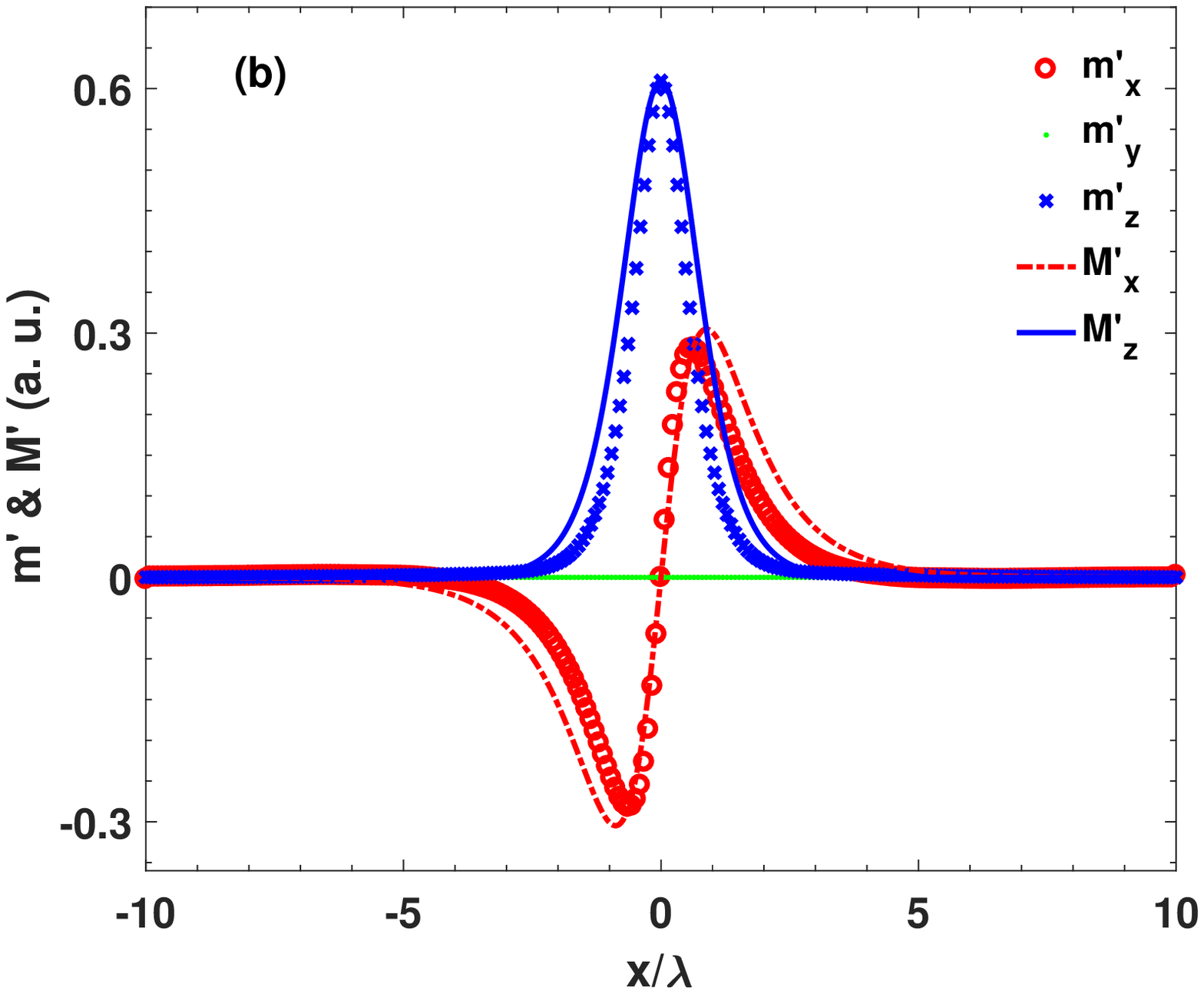}
\end{minipage}
\caption{Spatial derivative of the equilibrium itinerant magnetization distribution (\textbf{m}$'$) inside the DW region, with the DW width $\lambda k_F = 1$ (a) and $\lambda k_F = 70$ (b). The spatial derivative of the underlying $s$-$d$ exchange field (\textbf{M}$'$) is also displayed. The misalignment between \textbf{M}$'$ and \textbf{m}$'$ is mainly caused by the Rashba interaction.}
\label{dmag}
\end{figure}

\section{Outline of theory}
\label{theory}
We consider a special form of SOI in solids, the Rashba SOI \cite{Bychkov84} in a two dimensional (2D) electron system without spatial inversion symmetry. In a 2D electron gas, there could be electric field built up along the inversion symmetry breaking direction. In the rest frame of a moving electron, this static electric field is transformed into a magnetic field, and can influence the spin dynamics of moving electrons \cite{Gambardella}. Hence the Hamiltonian for itinerant electrons has the form \cite{Garate,Manchon,Matos-Abiague}
\begin{equation}
H = \frac{\textbf{p}^2}{2 m_e}  + \mu_B \bm{\sigma} \cdot \textbf{M} + \frac {\alpha_R} {\hbar} \bm{\sigma} \cdot (\textbf{p} \times \hat{z}),
\label{hamil}
\end{equation}
where $m_e$ is the electron mass, $\mu_B$ the Bohr magneton, and $\hbar$ the reduced Planck's constant. $\textbf{p}= -i \hbar \nabla$ is the momentum operator. $\alpha_R$ is the Rashba constant, which characterizes the broken inversion symmetry \cite{Bychkov84}. $\bm{\sigma} = \hat{x}\sigma_x + \hat{y}\sigma_y + \hat{z} \sigma_z$ is the vector Pauli matrix, which is also the electron spin operator if a multiplicative constant is ignored, with $\sigma_x$, $\sigma_y$ and $\sigma_z$ being the Pauli matrices. It is obvious from Eq. (\ref{hamil}) that the effective Rashba field is perpendicular to the symmetry breaking direction, which is $\hat {z}$ in the current case. The Hamiltonian (\ref{hamil}) describes the energy of conduction electrons in a solid, interacting through the $s$-$d$ exchange interaction with the localized electrons. For the purpose of illustrating the effect of the Rashba SOI on the STT, we only consider the spin dynamics of itinerant electrons dictated by the Hamiltonian (\ref{hamil}), while the local magnetic moments are assumed to be static, as described by the magnetization texture $\textbf{M}$. As our model Hamiltonian does not include the Coulomb interaction between electrons explicitly, there is no physical exchange interaction between electron spins. We use the magnetization texture $\textbf{M}$ to simulate the exchange interaction between conduction electrons.

The Rashba SOI's particular form in Eq. (\ref{hamil}) allows to define a covariant derivative operator \cite{kim13}, $\textbf{D} = \nabla + i m _e \alpha _R \hat {z} \times \bm {\sigma}/\hbar^2$. With the covariant derivative operator in place of the ordinary derivative operator, the Rashba SOI is absorbed into the kinetic energy term. The appearance of a position-independent emergent gauge potential in the covariant derivative signifies the chiral feature of the Rashba SOI, but in a different form. However, this chiral feature does not manifest itself in states with uniform magnetization distribution, except that the electronic energy band is spin-split and a global phase factor appears for the Bloch wave function. The situation changes for magnetization textures, as the gauge potential becomes position dependent if the magnetization vector varies in space, after applying a unitary transformation to make the magnetization parallel to the $z$ direction. Then the corresponding SU(2) magnetic field is nonzero and will affect electron's motion. It is this nonzero position and spin dependent magnetic field that gives rise to the chiral characteristics of the electron motion, and it is a general feature for any forms of SOI.

We will confine our discussion on STT solely to the Rashba SOI. But the Weyl SOI \cite{Kikuchi16} is related to the Rashba SOI by a unitary rotation in the spinor space, $U = \exp {(i \pi \sigma_z/4)}$, through the relation $\sigma_y p_y + \sigma_x p_x = U ^\dagger (\sigma_x p_y - \sigma_y p_x) U$. The Weyl form of SOI is compatible with a Bloch DW as the ground state. So if substitutions $- m _y \rightarrow m _x$, $m _x \rightarrow m _y$ and similar ones for other magnetic vectors are made, our results can be applied to the situation where the Weyl SOI is involved. The Dresselhaus SOI due to bulk inversion symmetry breaking \cite{Matos-Abiague}, $\sigma_x p_x - \sigma_y p_y$, differs from the Weyl SOI only by a transformation $y \rightarrow - y$. Since the magnetization textures considered by us vary only along the $x$ direction, the Hamiltonian (\ref{hamil}) is invariant under the inversion along the $y$ direction. Hence the effect of changing from the Weyl SOI to Dresselhaus SOI amounts to a substitution $m _y \rightarrow - m _y$. Given the above considerations, our conclusion about the STT nonadiabaticity and chirality should apply individually to all three types of SOI. Some systems possess both the Rashba SOI and the Dresselhaus SOI, such as the exemplar (001) GaAs/Fe interface discussed by Matos-Abiague and Rodr\'{\i}guez-Su\'{a}rez \cite{Matos-Abiague}. In such systems, our results cannot be applied directly and separate numerical investigations are needed. However, as our perturbation analysis (Sec. \ref{results}) will show, the chirality of $\alpha$ and $\beta$ and the algebraic decay of the nonadiabaticity are both derived from the coupling between the orbital and spin degrees of freedom. So as far as there is the presence of any forms of SOI, the coupling between spin and orbital motion will occur, and we stipulate that $\alpha$ and $\beta$ should be chiral and the decay of nonadiabaticity algebraic.

The magnetization texture considered is a N\'{e}el DW described by the unit magnetization vector $\hat {M} = \chi \hat{x} \mbox {sech} (x/\lambda) - q \hat{z} \tanh (x/\lambda)$, which is the renowned Walker profile \cite{Schryer74}. $\lambda = \sqrt {A/K}$ is the DW width that is determined by the material specific exchange and anisotropy constants $A$ and $K$. $\hat {x}$ and $\hat {z}$ are unit vectors pointing along the $x$ and $z$ directions, respectively. The charge $q$ and the chirality $\chi$ \cite{Braun} of the DW are topological numbers to quantify its topological characteristics. Our definition of the DW chirality differs from that in Ref. [\onlinecite{Braun}] by a factor $q$, so the product $q \chi$ is actually the DW chirality defined in Ref. [\onlinecite{Braun}]. We assume then the current is flowing along the $x$ direction, and the electrons are moving in the 2D $xy$ plane. The eigenvalue problem corresponding to the Pauli-Schr\"{o}dinger equation with the Hamiltonian (\ref{hamil}) and the Walker magnetization profile as given above is difficult to solve analytically. We adopt a scattering matrix method to numerically solve the eigenvalue problem \cite{Xiao06,Xia06,Zwierzycki08}. The physical picture behind such a scattering method is simple: We inject plane waves which are the solutions to a uniform magnetization distribution from both $\pm \infty$, then let them evolve according to the Hamiltonian (\ref{hamil}) and match the evolved waves at the DW center, by requiring the continuity of wave functions and their first order derivatives.

With the wave functions thus obtained, the ground state magnetization distribution can be expressed as an integral over the Fermi sphere in 2D,
\begin{equation}
\textbf{m} (x) = \int \frac{d^2 k} {(2\pi)^2} \psi^ \dagger _\textbf{k} (x) \bm{\sigma} \psi _\textbf{k} (x),
\label{mk}
\end{equation}
where $\textbf{k}$ is a Bloch wave vector in the momentum space. We then employ a semiclassical approach with relaxation time approximation \cite{Ashcroft} to calculate the STT in the current-carrying state,
\begin{equation}
\bm{\tau} (x) = - \frac{e E \tau_0} {(2\pi)^2\hbar} \oint d\varphi\, k_x\, \textbf{Q}' _\textbf{k} (x),
\label{tauk}
\end{equation}
where $\tau_0$ is the relaxation time constant, $e$ is the electron charge, and $\varphi$ the angle of the wave vector relative to the $x$ axis. $\textbf{Q}' = d \textbf{Q}/dx$ denotes the $x$ derivative of the spin current density $\textbf{Q}$. Since we only consider the zero temperature situation, the integration is confined to the Fermi surface in 2D, which is a circle. The spin current density $\textbf{Q}$ is defined as
\begin{equation}
\textbf{Q} = - 2 \Im \left[ \psi ^ \dagger _\textbf{k} (x) \bm{\sigma} \psi' _\textbf{k} (x)\right] + \hat {y} k _R \psi ^ \dagger _\textbf{k} (x) \psi _\textbf{k} (x).
\label{qk}
\end{equation}
$k _R$ is an effective wave number proportional to the Rashba coupling constant, $\hbar^2 k_R /2 m_e = \alpha_R$, and $\hat {y}$ is a unit vector along the $y$ direction. The full spin current density is a seocnd-rank tensor, with one index in the spinor space and the other index in the coordinate space. The form $\textbf{Q}$ presented here is actually the $x$ component of the full spin current density tensor. As the spin current density tensor depends on the $x$ coordinate, in accordance with the magnetization profile considered here, only the $x$ component contributes to the calculation of the STT, as shown in Eq. (\ref{qk}), and we do not need the full tensor. Further details and particulars on numerical wave function and torque calculation can be found in Ref. [\onlinecite{Wang19}].
\begin{figure}\centering
\begin{minipage}[c]{0.45\linewidth}
\includegraphics[width=\linewidth]{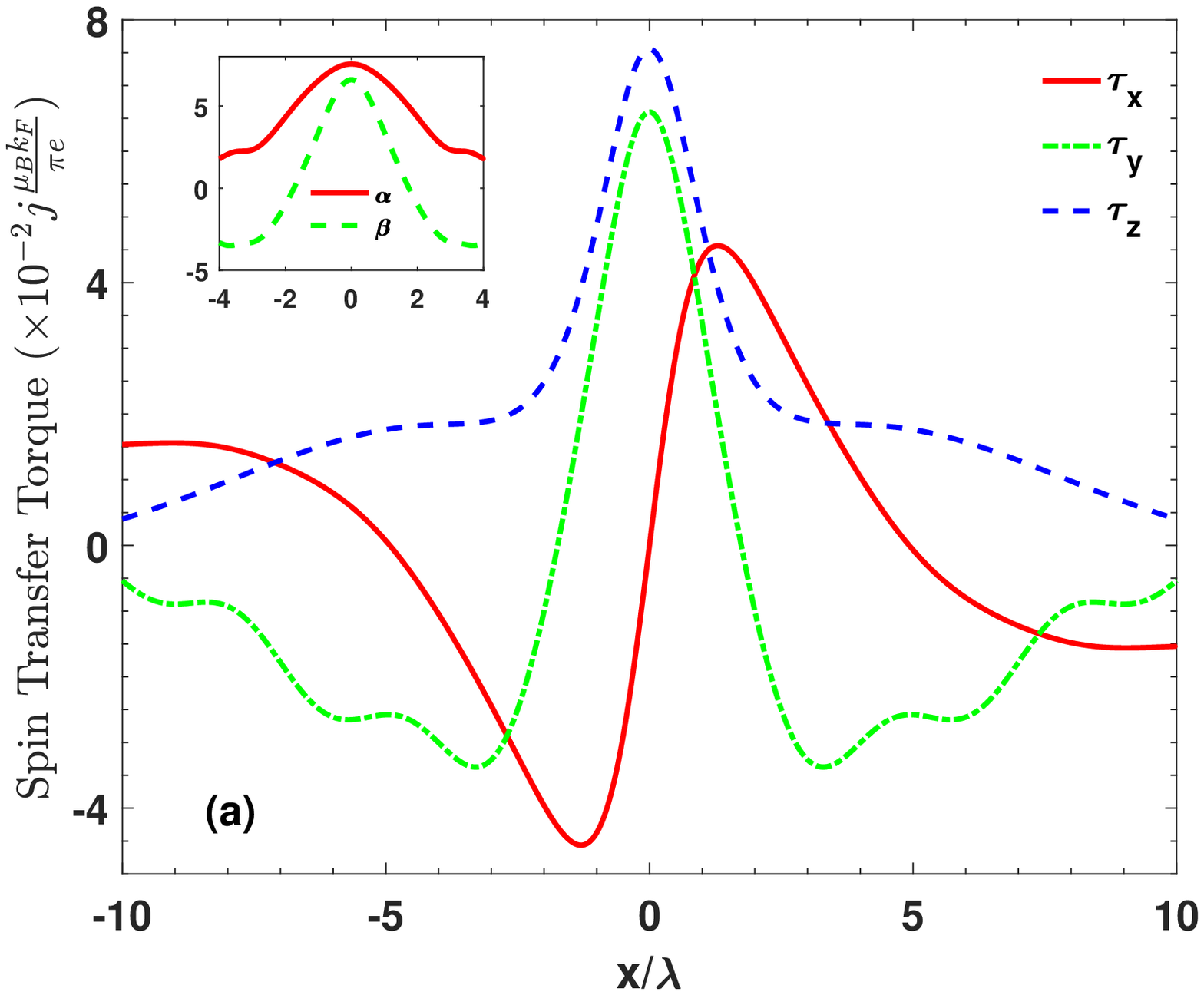}
\end{minipage}
\hfill
\begin{minipage}[c]{0.45\linewidth}
\includegraphics[width=\linewidth]{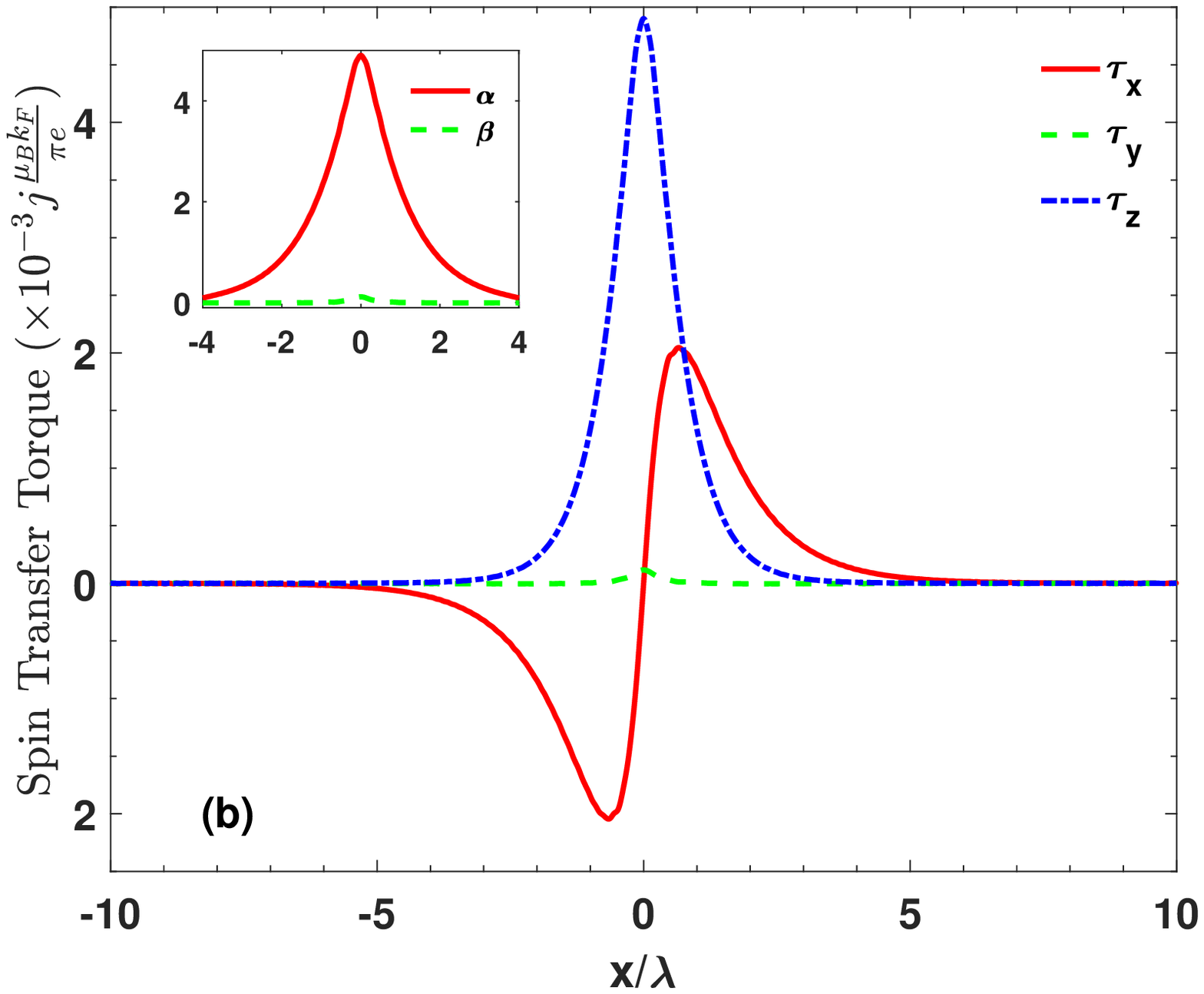}
\end{minipage}
\caption{Numerically calculated STT, $\bm {\tau}$, for $\lambda k_F = 1$ (a) and $\lambda k_F = 70$ (b). The corresponding decomposition coefficients $\alpha$ and $\beta$ are displayed in the insets. The observable oscillation for $\lambda k_F = 1$ is due to the quantum-confinement effect induced by the presence of the short DW. As the DW width is increased, the quantum oscillation is smoothed out, as shown in (b). The unit for the 2D STT is $j \mu _B k _F/\pi e$, where $j$ is the current density for the free electron gas and $\mu _B$ is the Bohr magneton. As we use only unit vectors for the decomposition of the STT, Eq. (\ref{sttiti}), the decomposition coefficients $\alpha$ and $\beta$ have the same unit as that for the STT.}
\label{sttw}
\end{figure}

\section{Numerical results}
\label{results}
For the convenience of numerical calculation and a direct comparison between different quantities, we convert the local exchange field $M$ through relation $\hbar^2 k_B^2/2 m_e = \mu_B M$ into an effective wave number, as we already did in the preceding section for the Rashba coupling constant $\alpha_R$. We then measure them in terms of the Fermi wave number $k_F$ for the free electron gas, the dynamics of which is governed only by the first term in the Hamiltonian (\ref{hamil}). In our following numerical results, we use the values $k_B/k_F = 0.4$ and $k_R/k_F = 0.1$ unless stated otherwise. The numerically obtained spatial derivative of the equilibrium itinerant magnetization distribution $\textbf{m}'$ is given in Fig. 1, together with the background local magnetization's spatial derivative, for two typical DW width values, $\lambda k_F = 1$ and $\lambda k_F = 70$. As the transition from the nonadiabatic to adiabatic behaviour is defined by the critical DW width \cite{Xiao06} $\lambda_c k_F = k_F^2/k_B^2 = 6.25$ for $k_B/k_F = 0.4$, the value we used to introduce the exchange interaction for itinerant magnetization, $\lambda k_F = 1$ corresponds to the extremely nonadiabatic situation and $\lambda k_F = 70$ the adiabatic one. In the extremely nonadiabatic limit, the spatial variation of the itinerant magnetization is distributed over the whole simulated region and the misalignment between $\textbf{m}'$ and $\textbf{M}'$ is not negligible. In the adiabatic limit, however, only the spatial variation is concentrated to the DW center region and the misalignment between $\textbf{m}'$ and $\textbf{M}'$ is still there. The oscillation of the itinerant magnetization observable for short DWs is attributable to the quantum-confinement effect induced by the presence of the local magnetization profile $\textbf{M}$. It actually decays away very quickly: For $\lambda k_F = 3$ (not shown here), the quantum oscillation is already not discernable. The misalignment between $\textbf{m}'$ and $\textbf{M}'$ can be traced back to the finite Rashba coupling present in our calculation. As was discussed in Ref. [\onlinecite{Wang19}], the definite parity for the itinerant magnetization components arises due to the parity-time, or particle-hole, symmetry of the Hamiltonian (\ref{hamil}). It is interesting to note that the same symmetry was also observed for magnons inside DWs \cite{Wang17}.
\begin{figure}\centering
\begin{minipage}[c]{0.45\linewidth}
\includegraphics[width=\linewidth]{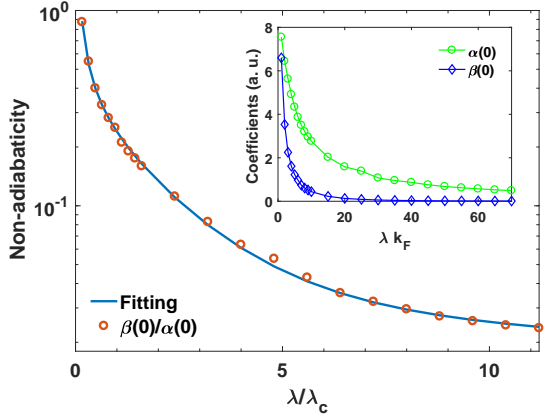}
\end{minipage}
\caption{Nonadiabaticity $\beta (0)/\alpha(0)$ as a function of the DW width $\lambda$, with individual coefficients $\alpha(0)$ and $\beta(0)$ shown in the inset. Obviously, the decay of the nonadiabaticity is not simply exponential. The solid line in the main panel is the result fitted according to Eq. (\ref{nonadiab}), while those in the inset are merely guides to the eye.}
\label{adia}
\end{figure}

A qualitatively similar behaviour is observed for the STT, as shown in Fig. \ref{sttw}. In the extremely nonadiabatic limit, $\lambda k_F = 1$, both the adiabatic and nonadiabatic components of the STT contribute, manifested by the finite value of both $\alpha$ and $\beta$. As we consider only the dynamics of the itinerant magnetization $\textbf{m}$, the actual decomposition of the STT is
\begin{equation}
\bm {\tau} = \alpha\, \hat{m}' + \beta\, \hat {m} \times \hat{m}',
\label{sttiti}
\end{equation}
using the unit vector along the direction of the derivative of the itinerant magnetization vector $\hat {m}' = \textbf{m}'/m'$ instead of the derivative of the local magnetization vector. The magnitude of $\alpha$ and $\beta$ is comparable to each other, signifying the significant contribution of the nonadiabatic STT. As the DW width is increased, both $\alpha$ and $\beta$ decrease in magnitude, with the relative importance of the nonadiabatic component decreasing, too. As the nonadiabatic STT acts as an effective field-like torque in the Landau-Lifshitz-Gilbert equation \cite{llg} describing the magnetization dynamics phenomenologically, the relative importance of the nonadiabatic STT warrants more investigation.

To measure the relative importance of the nonadiabatic STT, we follow Ref. [\onlinecite{Xiao06}] to define the nonadiabaticity of the STT as the ratio between the nonadiabatic and adiabatic coefficients at the DW center, $\beta(0)/\alpha(0)$, where the spatial variation of the local magnetization is maximized. The nonadiabaticity thus defined is plotted in Fig. \ref{adia}, along with the individual coefficients as a function of the DW width. Surprisingly, the behaviour of the nonadiabatitity in the presence of the Rashba SO coupling shows a much slower decay in the DW width, as compared to the case without the Rashba coupling. Although the nonadiabaticity in Fig. \ref{adia} decays away as a whole, its decay is not exponential anymore, but resembling more like a power-law decay, which is also in contrast to the previously obtained oscillatory behaviour through analytical treatment of a semiclassical kinetic equation \cite{Bohlens10} or the unusual linear increase scaling extracted from density functional theory calculation \cite{Yuan16}.
\begin{figure}\centering
\begin{minipage}[c]{0.45\linewidth}
\includegraphics[width=\linewidth]{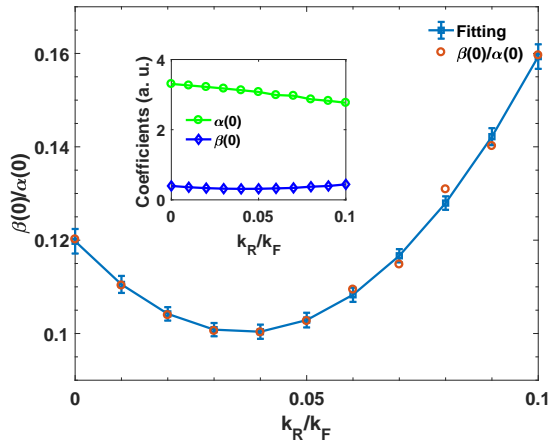}
\end{minipage}
\caption{Dependence of the nonadiabaticity on the Rashba coupling strength for $\lambda k_F = 10$, with $\alpha(0)$ and $\beta(0)$ shown in the inset. The solid line in the main panel is fitted to a parabola, while those in the inset are only guides to the eye.}
\label{alpha}
\end{figure}

The behaviour of the STT nonadiabaticity can be understood qualitatively by a perturbation analysis of the Pauli-Schr\"{o}dinger equation. Using a unitary transformation $H _ \alpha = U^ \dagger H U$ with $U = \exp {(-i \alpha \sigma_y/2)}$, we can transform the Hamiltonian (\ref{hamil}) into a different form in the spinor space, $H _\alpha = - [\partial_x - i (k_ R + \alpha') \sigma_y/2 ]^2 + \xi_z \sigma_z + k_ R ^2/4$, which is measured in units of the Fermi energy of the free electron gas, in accordance with we did for the wave numbers. The positive local effective exchange field $\xi_z$ is given by the relation $\xi_z^2 = k_B^4 + k_ R^2 k_y^2 + 2 k_ R k_y k_B^2 \sin \theta$, depends on both position and momentum. The rotation angle $\alpha$ is related to the magnetization angle $\theta$ and the electron's transverse momentum $k_y$ \cite{Wang19}, $\tan \alpha = \tan \theta + k_ R k_y/k_B^2 \cos \theta$. The Hamiltonian $H _\alpha$ can be expanded into the sum of a position-independent, unperturbed part $H _0$ and a position-dependent, perturbation potential $V$, $H _\alpha = H_0 + V$. The unperturbed Hamiltonian is given by $H_0 = - \partial_x ^2 + \xi_0 \sigma_z + i k_ R \sigma_y \partial_x$, and the perturbation is $V = k_ R \alpha'/2 + \alpha'^2/4 + i \sigma_y ( \alpha' \partial_x + \alpha''/2 ) + (\xi_z - \xi_0) \sigma_z$, with a constant exchange field $\xi_0 = \sqrt {k_B^4 + k_ R^2 k_y^2}$. In the adiabatic ($\lambda \rightarrow \infty$) and weak Rashba coupling ($\alpha_R \rightarrow 0$) limit, the position dependent part $V$ can be treated as a perturbation to the position independent Hamiltonian $H _0$. In accordance with our numerical scattering approach to solve the eigenvalue problem corresponding to the Hamiltonian (\ref{hamil}), what appears in our perturbation analysis is actually the potential in momentum space, with incoming momentum $k _i$ and scattered momentum $k _f$ along the $x$ direction,
\begin{eqnarray}
V(k_f, k_i) &=& \frac {p \, \mbox {csch} p} {8 \pi \lambda} - \chi \frac {k_ R k_y} {k_B^2} \frac {\pi^2 + 4 p^2} {16 \pi^2 \lambda} \mbox {sech} p + q \chi\frac {k_ R} {8} \mbox {sech} p \nonumber\\
&-& q \chi \frac {k_s} {8} \left(\mbox {sech} p - 2 \chi \frac {k_ R k_y} {\pi k_B^2} p \, \mbox {csch} p \right) \sigma_y + \chi\frac {\lambda k_ R} {4} k_y \sigma_z \mbox {sech} p.
\label{vk}
\end{eqnarray}
$p = (k_f - k_i) \lambda \pi/2$ is the normalized momentum transfer and $k_s = k_f + k_i$ the sum of the injected and scattered momenta. As the anti-Hermitian part of $V(k_f, k_i)$ involves the digamma function and is very complicated, we only show the Hermitian part in Eq. (\ref{vk}). Correction to wave functions can be obtained by considering the scattering of the zeroth-order wave functions by the potential $V(k_f, k_i)$. With the obtained perturbative wave functions, we can proceed to calculate the magnetization and the STT using Eqs. (\ref{mk}), (\ref{tauk}) and (\ref{qk}), then decompose the STT according to Eq. (\ref{sttiti}) to get the coefficients $\alpha$ and $\beta$. Further calculation details can be found in Ref. [\onlinecite{Wang19}], so we only give the final results in the following.

To the lowest order, the equilibrium itinerant magnetization $\textbf{m}$ is everywhere parallel to the local magnetization $\textbf{M}$, $\textbf{m} \propto \textbf{M}$. Correspondingly, their spatial derivatives are also parallel to each other, $\textbf{m}' \propto \textbf{M}' = (\hat {x} \cos \theta - \hat {z} \sin \theta) \theta'$. Recalling the definition of the magnetization angle $\theta$, we can immediately calculate its spatial derivative, $\lambda \theta' = q \chi \mbox {sech} (x/\lambda)$, which is proportional to the product of the charge and the chirality. The phenomenological decomposition of the STT (\ref{sttiti}) then has the form $\bm {\tau} = \alpha q \chi (\hat {x} \cos \theta - \hat {z} \sin \theta) + \hat {y} \beta q \chi$. The lowest order perturbation result for the STT gives $\bm {\tau} \propto \textbf{m}' $, where only the adiabatic component contributes. According to this expression, the adiabatic coefficient at the DW center $\alpha(0)$ should scale inversely proportional to the DW width $\lambda$, $\alpha(0) \propto 1/\lambda$. Including the first order correction to wave functions, a finite nonadiabatic STT at the DW center emerges, which is proportional to
\begin{equation}
q \chi \left( \frac {a} {\lambda^2} + q \chi  k_R \frac {b + c e^ {- \gamma \lambda}} {\lambda}\right),
\label{beta}
\end{equation}
where $a$, $b$, $c$ and $\gamma$ are constants determined by the effective exchange splitting $k_B$ and possibly the Rashba coupling strength $k_R$. The exponential and inverse power law terms are brought about by the terms with non-zero and zero momentum transfers in the momentum-space effective potential \cite{Dugaev02,Wang19}. Using this result for the coefficient $q \chi \beta(0)$, the nonadiabaticity has the form
\begin{equation}
\frac {\beta(0)} {\alpha(0)} =\frac {a} {\lambda} + q \chi k_R (b + c e^ {- \gamma \lambda}).
\label{nonadiab}
\end{equation}
The corresponding fit to the expression (\ref{nonadiab}) is displayed in Fig. \ref{adia}. The agreement of the fit to the numerical result is satisfactory.
\begin{figure}\centering
\begin{minipage}[c]{0.45\linewidth}
\includegraphics[width=\linewidth]{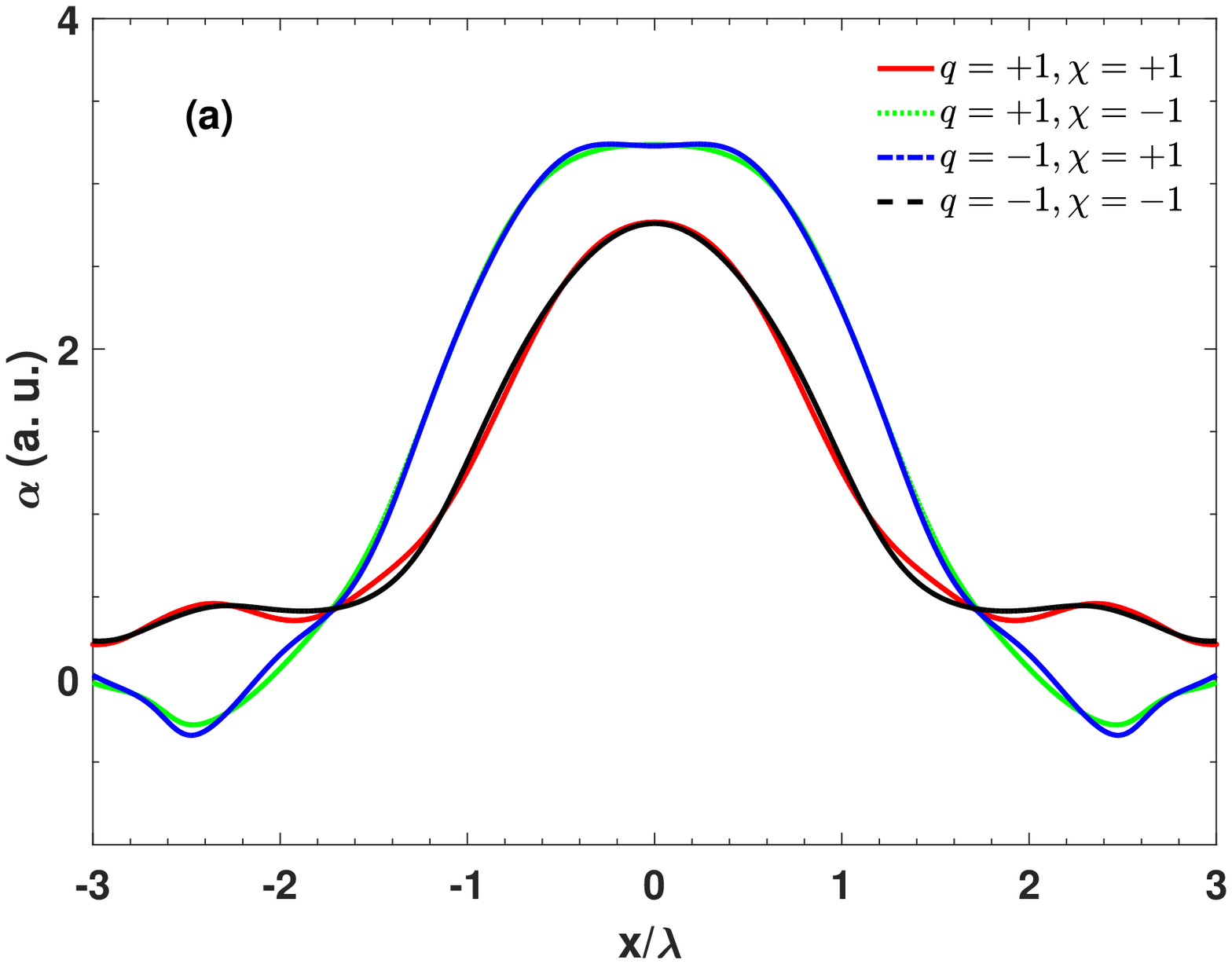}
\vspace{0.01\textheight}
\end{minipage}
\hfill
\begin{minipage}[c]{0.45\linewidth}
\includegraphics[width=\linewidth]{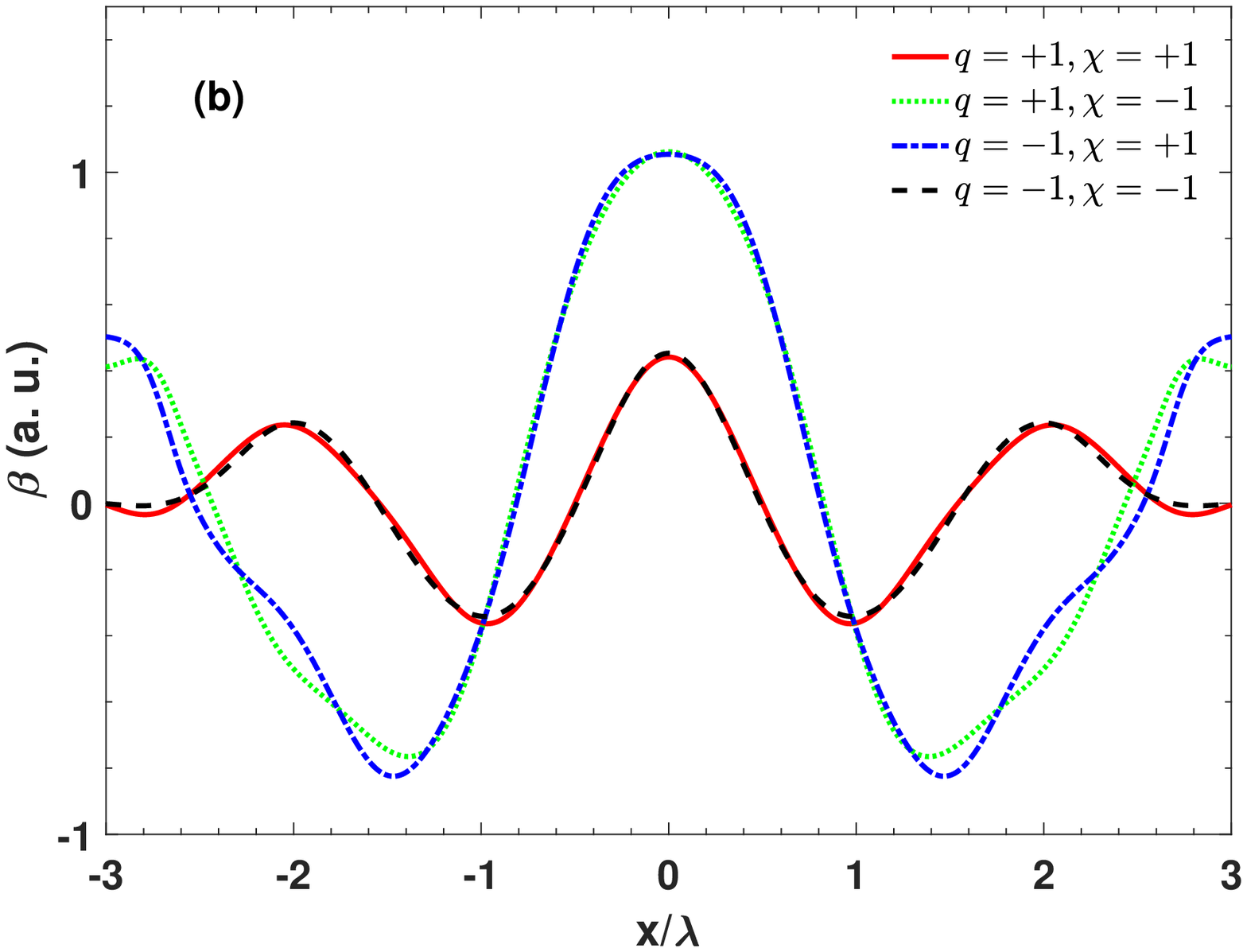}
\vspace{0.01\textheight}
\end{minipage}
\begin{minipage}[c]{0.45\linewidth}
\includegraphics[width=\linewidth]{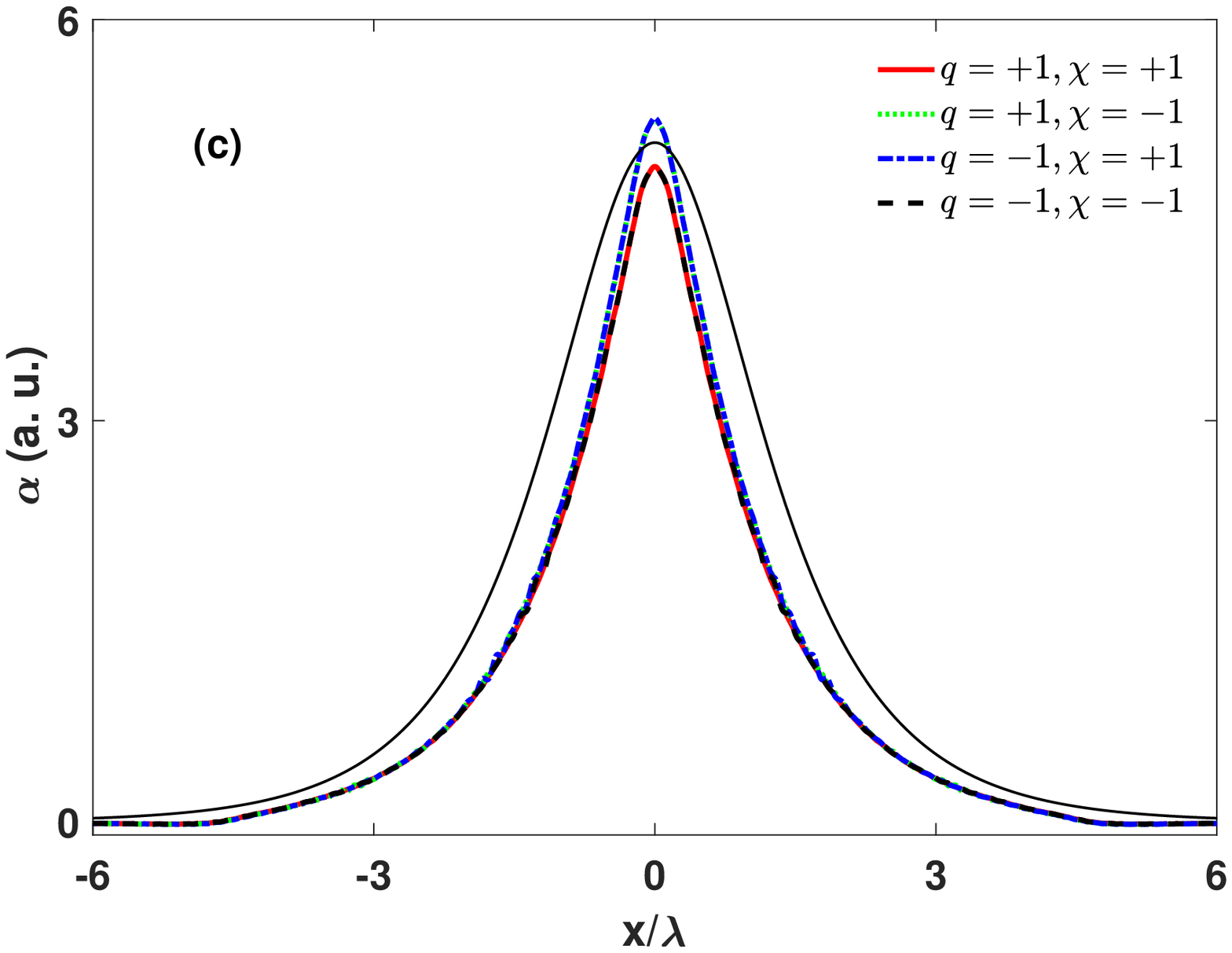}
\end{minipage}
\hfill
\begin{minipage}[c]{0.45\linewidth}
\includegraphics[width=\linewidth]{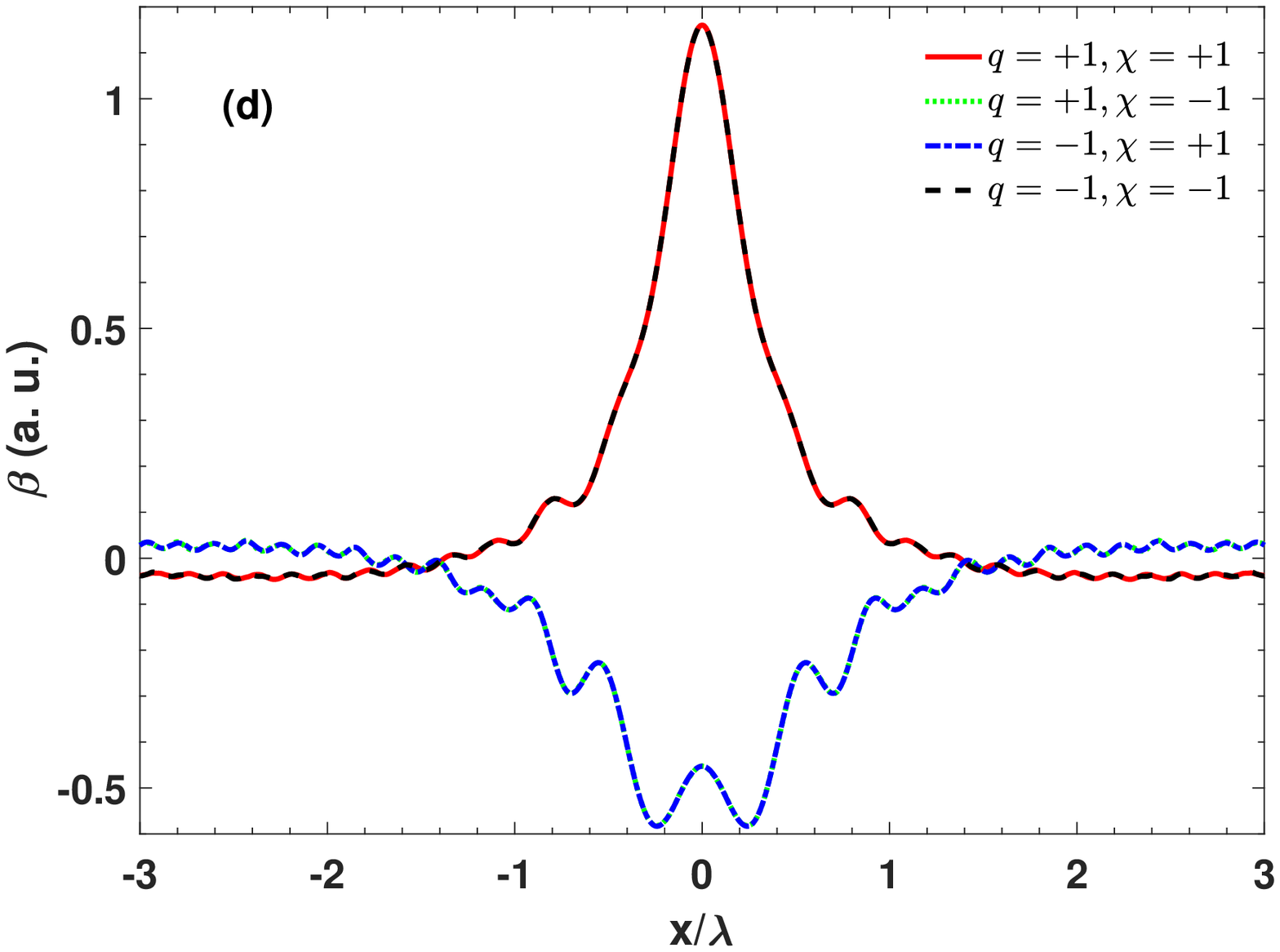}
\end{minipage}
\caption{Chiral decomposition coefficients $\alpha$ and $\beta$ for different combinations of the DW charge $q$ and chirality $\chi$ with $\lambda k_F = 10$ (a, b) and $\lambda k_F = 70$ (c, d). It is evident that both $\alpha$ and $\beta$ are almost completely determined only by the product $q \chi$, rather than their separate values. The thin solid line shown in (c) is proportional to the $x$ derivative of the magnetization angle, $\theta'$. In the adiabatic and vanishing Rashba SOI limit, $\alpha$ should be proportional to $\theta'$. The deviation from this asymptotic behaviour can result in the appearance of terms higher order in the derivative $\theta'$, in the expansion of the STT in terms of the variation of magnetization in space.}
\label{sttq}
\end{figure}

The first order perturbative result can only be used to understand the DW width dependence of the nonadiabaticity of the STT in the adiabatic limit. In fact, there could be higher-order terms contributing to the multiplicative coefficients $a$, $b$ and $c$ of the exponential and power functions of $\lambda$. Especially for short DWs, the nonadiabacity in Fig. \ref{alpha} shows significant nonlinear dependence on the Rashba coupling constant, although the nonlinear behaviour of both individual coefficients exhibit mild impact of higher order terms, cf. insets to Fig. \ref{alpha}.

Finally, we would like to check that the coefficients $\alpha$ and $\beta$ are chiral under the influence of the Rashba SOI. Due to the fact that the unit vector $\hat{m}'$ is proportional to the product $q \chi$, this geometrical factor has to be factored out in the consideration of the chirality of $\alpha$ and $\beta$, as explicitly shown in Eq. (\ref{beta}), where the expression for $\beta$ is given in the parentheses. As is obvious from the expression for $\beta$ in Eq. (\ref{beta}), in addition to a term independent of the topological features of the DW, the term proportional to the Rashba interaction depends on the product of the DW's charge and chirality, giving rise to a dynamical dependence on the DW topology. Although this behaviour is in stark contrast to that of the nonadiabatic RSOT which is solely determined by the product of charge and chirality \cite{Wang20}, it is actually borne out by the numerical results, as shown in Fig. \ref{sttq}. For short DWs, the Rashba contribution to $\beta$ is smaller compared to the contribution to the STT arising from the spatial variation of the magnetization, and the nonadiabatic STT is dominated by the spatial variation of the magnetization. In the adiabatic limit, the Rashba contribution becomes important, so the nonadiabatic coefficient $\beta$ shows the characteristic $q \chi$ behaviour (Fig. \ref{sttq} (d)), although there is still some deviation from the perfect $q \chi$ behaviour due to the contribution from the spatial variation of the magnetization. A similar behaviour can be observed for the adiabatic coefficient $\alpha$ in Fig. \ref{sttq}. The dependence on the product $q \chi$ makes both $\alpha$ and $\beta$ chiral, reminiscent of the chiral nature of the Rashba SOI and confirming our expectation that the coefficients $\alpha$ and $\beta$ should be chiral.
\section{Conclusion}
\label{conclusion}
To conclude, we have investigated the DW width scaling of STT in magnetic DWs with a sizable Rashba SOI, which is originated from the broken inversion symmetry at ferromagnet/heavy metal interfaces. In the conventional case where only the spatial variation of magnetization is responsible for the emergence of the nonadiabatic STT, an exponential decay for the nonadiabaticity, which is used to measure the relative importance of the nonadiabatic to the adiabatic torques, was observed. In contrast, in DWs with Rashba SOI, the decay of the nonadiabaticity is algebraic, much slower than an exponential behaviour. Due to the presence of the finite Rashba SOI, both the adiabatic and the nonadiabatic decomposition coefficients for the STT exhibit chiral dependence on the topology of the underlying DW, especially in the adiabatic limit where the Rashba contribution is dominant. This chiral feature of the STT decomposition coefficients is absent in systems without Rashba SOI.

\section*{Acknowledgements}
We would like to express our gratitude to Prof. Jiang Xiao for valuable comments and discussions, especially for bringing us to the topic of STT in magnetic DWs with Rashba spin orbit interaction and sharing his code on STT simulation. Y. Z. acknowledges the support by the President's Fund of CUHKSZ, Longgang Key Laboratory of Applied Spintronics, National Natural Science Foundation of China (Grants No. 11974298, No. 61961136006), Shenzhen Fundamental Research Fund (Grant No. JCYJ20170410171958839), and Shenzhen Peacock Group Plan (Grant No. KQTD20180413181702403).


\begin{thebibliography}{99}
\bibitem{Berger}
L. Berger, J. Appl. Phys. 49, 2156 (1978); Phys. Rev. B 54, 9353 (1996).

\bibitem{Slonczewski96}
J. Slonczewski, J. Magn. Magn. Mater. 159, L1 (1996).

\bibitem{Kirilyuk}
A. Kirilyuk, A. V. Kimel, and Th. Rasing, Rev. Mod. Phys. 82, 2731 (2010).

\bibitem{Ohno}
H. Ohno, D. Chiba, F. Matsukura, T. Omiya, E. Abe, T. Dietl, Y. Ohno, and K. Ohtani, Nature (London) 408, 944 (2000).

\bibitem{Dreher12}
L. Dreher, M. Weiler, M. Pernpeintner, H. Huebl, R. Gross, M. S. Brandt, and S. T. B. Goennenwein, Phys. Rev. B 86, 134415 (2012); Erratum: \textit{ibid.} 98, 099901 (2018).

\bibitem{Obata}
K. Obata and G. Tatara, Phys. Rev. B 77, 214429 (2008).

\bibitem{Manchon}
A. Manchon and S. Zhang, Phys. Rev. B 78, 212405 (2008); \textit{ibid.} 79, 094422 (2009).

\bibitem{Garate}
I. Garate and A. H. MacDonald, Phys. Rev. B 80, 134403 (2009).

\bibitem{Matos-Abiague}
A. Matos-Abiague and R. L. Rodr\'{\i}guez-Su\'{a}rez, Phys. Rev. B 80, 094424 (2009).

\bibitem{Wang12}
X. Wang and A. Manchon, Phys. Rev. Lett. 108, 117201 (2012).

\bibitem{she}
J. E. Hirsch, Phys. Rev. Lett. 83, 1834 (1999); S. F. Zhang, Phys. Rev. Lett. 85, 393 (2000); J. Sinova, D. Culcer, Q. Niu, N. A. Sinitsyn, T. Jungwirth, and A. H. MacDonald, Phys. Rev. Lett. 92, 126603 (2004).

\bibitem{Bazaliy98}
Ya. B. Bazaliy, B. A. Jones, and S.-C. Zhang, Phys. Rev. B 57, R3213 (1998).

\bibitem{Li04}
Z. Li and S. Zhang, Phys. Rev. Lett. 93, 127204 (2004).

\bibitem{Thiaville05}
A. Thiaville, Y. Nakatani, J. Miltat, and Y. Suzuki, Europhys. Lett. 69,
990 (2005).

\bibitem{Xiao06}
J. Xiao, A. Zangwill and M. D. Stiles, Phys. Rev. B 73, 054428 (2006).

\bibitem{Gambardella}
P. Gambardella and I. M. Miron, Philos. Trans. R. Soc. London, Ser. A 369, 3175 (2011).

\bibitem{kim13}
K.-W. Kim, H.-W. Lee, K.-J. Lee, and M. D. Stiles, Phys. Rev. Lett. 111, 216601 (2013).

\bibitem{Wang20}
D. Wang and Y. Zhou, Phys. Rev. B 101, 020410(R) (2020).

\bibitem{Jue}
E. Ju\'{e}, C. K. Safeer, M. Drouard, A. Lopez, P. Balint, L. Buda-Prejbeanu, O. Boulle, S. Auffret, A. Schuhl, A. Manchon, I. M. Miron, and G. Gaudin, Nat. Mater. 15, 272 (2016).

\bibitem{Akosa}
C. A. Akosa, I. M. Miron, G. Gaudin, and A. Manchon, Phys. Rev. B 93, 214429 (2016).

\bibitem{Freimuth17}
F. Freimuth, S. Bl\"{u}gel, and Y. Mokrousov, Phys. Rev. B 96, 104418 (2017).

\bibitem{Bychkov84}
Yu. A. Bychkov and E. I. Rashba, JETP Lett. 39, 78 (1984).

\bibitem{Kikuchi16}
T. Kikuchi, T. Koretsune, R. Arita, and G. Tatara, Phys. Rev. Lett. 116, 247201 (2016).

\bibitem{Schryer74}
N. L. Schryer and L. R. Walker, J. Appl. Phys. 45, 5406 (1974).

\bibitem{Braun}
H.-B. Braun, Adv. Phys. 61, 1 (2012).

\bibitem{Xia06}
K. Xia, M. Zwierzycki, M. Talanana, P. J. Kelly, and G. E. W. Bauer, Phys. Rev. B 73, 064420 (2006).

\bibitem{Zwierzycki08}
M. Zwierzycki, P. A. Khomyakov, A. A. Starikov, K. Xia, M. Talanana, P. X. Xu, V. M. Karpan, I. Marushchenko, I. Turek, G. E. W. Bauer, G. Brocks, and P. J. Kelly, Phys. Stat. Sol. B 245, 623 (2008).

\bibitem{Ashcroft}
N. W. Ashcroft and N. D. Mermin, \textit{Solid State Physics}, World Publishing Corporation, Beijing, 2004.

\bibitem{Wang19}
D. Wang and Y. Zhou, J. Magn. Magn. Mater. 493, 165694 (2020).

\bibitem{Wang17}
D. Wang, Y. Zhou, Z.-X. Li, Y. Nie, X.-G. Wang, and G.-H. Guo, IEEE Trans. Magn. 53, 1300110 (2017).

\bibitem{llg}
L. D. Landau, E. M. Lifshitz, and L. P. Pitaevski, Statistical Physics, 3rd ed. (Pergamon, Oxford), Part 2, 1980; T. L. Gilbert, IEEE Trans. Magn. 40, 3443 (2004).

\bibitem{Bohlens10}
S. Bohlens and D. Pfannkuche, Phys. Rev. Lett. 105, 177201 (2010).

\bibitem{Yuan16}
Z. Yuan and P. J. Kelly, Phys. Rev. B. 93, 224415 (2016).

\bibitem{Dugaev02}
V. K. Dugaev, J. Barna\'{s}, A. {\L}usakowski, and {\L}. A. Turski, Phys. Rev. B 65, 224419 (2002).
\end{thebibliography}
\end{document}